# Region-specific Dictionary Learning-based Low-dose Thoracic CT Reconstruction

Qiong Xu, Jeff Wang, Hiroki Shirato, and Lei Xing

*Abstract*—This paper presents a dictionary learning-based method with region-specific image patches to maximize the utility of the powerful sparse data processing technique for CT image reconstruction. Considering heterogeneous distributions of image features and noise in CT, region-specific customization of dictionaries is utilized in iterative reconstruction. Thoracic CT images are partitioned into several regions according to their structural and noise characteristics. Dictionaries specific to each region are then learned from the segmented thoracic CT images and applied to subsequent image reconstruction of the region. Parameters for dictionary learning and sparse representation are determined according to the structural and noise properties of each region. The proposed method results in better performance than the conventional reconstruction based on a single dictionary in recovering structures and suppressing noise in both simulation and human CT imaging. Quantitatively, the simulation study shows maximum improvement of image quality for the whole thorax can achieve 4.88% and 11.1% in terms of the Structure-SIMilarity (SSIM) and Root-Mean-Square Error (RMSE) indices, respectively. For human imaging data, it is found that the structures in the lungs and heart can be better recovered, while simultaneously decreasing noise around the vertebra effectively. The proposed strategy takes into account inherent regional differences inside of the reconstructed object and leads to improved images. The method can be readily extended to CT imaging of other anatomical regions and other applications.

*Index Terms*—Low-dose CT, thoracic CT, dictionary learning, iterative reconstruction, regional regularization.

The authors would like to thank Drs. Hengyong Yu, Jiang Hiesh and Ge Wang for the rights of using the real data. This work was supported in part by NIH (5R01 CA176553).

Q. Xu and L. Xing are with the Department of Radiation Oncology, School of Medicine, Stanford University, Stanford, CA 94305, USA, and also with the Global Station for Quantum Medical Science and Engineering, Global Institution for Collaborative Research and Education (GI-CoRE), Hokkaido University, Sapporo 060-8648, Japan.

J. Wang is with the Global Station for Quantum Medical Science and Engineering, Global Institution for Collaborative Research and Education (GI-CoRE), Hokkaido University, Sapporo 060-8648, Japan.

H. Shirato is with the Faculty of Radiation Medicine, Hokkaido University, Sapporo 060-8648, Japan, and also with the Global Station for Quantum Medical Science and Engineering, Global Institution for Collaborative Research and Education (GI-CoRE), Hokkaido University, Sapporo 060-8648, Japan.

## I. Introduction

CT is widely used in diagnosis of diseases in the thoracic region, such as lung tumor and cardiac disease. It is also indispensable for treatment planning, interventional guidance, and therapeutic assessment. However, X-ray radiation dose is an important concern since ionizing radiation can increase the risk of cancer and other genetic diseases. Compared to a simple chest radiograph, the dose of thoracic CT is greatly increased. Development of low-dose CT has thus been an active area of research in recent years [1-5].

Many methods have been proposed to improve quality of low-dose CT imaging. The approaches can be classified into three categories. The first attempts to denoise the projection data by various filtering techniques and then perform the conventional reconstruction [6-9]. This kind of methods may introduce new artifacts caused by the inconsistency of filtered projection data. The second performs denoising in the image domain [10-15]. Some well-known denoising techniques, such as dictionary learning, neural networks, and deep learning, have been used. However, because the methods act on images reconstructed by conventional methods, such as FBP or FDK, some information lost in the reconstruction step is difficult to recover. The last category of methods combines the projection and image domains to perform iterative reconstruction [16-30]. These methods consider both data fidelity of the projections and prior information of the system in recovering the images. Although iterative reconstruction is time consuming, it can achieve superior image quality. Dictionary learning-based iterative reconstruction is one of the representative methods in this last category [27].

The assumption of the dictionary learning technique is that an image patch in CT can be expressed as a linear combination of a few basic image patch primitives. Given a set of training image patches, a smaller set of basic image patches can be obtained by an optimization procedure to sparsely represent each and every image patch in the training set [31-33]. This basic image patch set is called a dictionary and each patch in the dictionary is called an atom. The dictionary contains the primary structural features of the training set. The underlying assumption in applying the dictionary learning technique to CT reconstruction is that the patches extracted from the reconstructed image can be sparsely represented by the dictionary [27]. Therein, the dictionary learning-based CT reconstruction aims to minimize a cost function consisting of the projection data fidelity, the dictionary-based representation error, and representation sparsity, by an iterative optimization process. For dictionary-based reconstruction to perform well, the dictionary should have as many similar structures as possible, while containing as few irrelevant structures as possible.

To illustrate the problem that region-specific dictionary learning method attempts to solve, let us consider the single dictionary learning-based CT reconstruction method [27]. When training a dictionary from the patch set extracted from the whole thoracic image, the patches extracted from the lung region usually contain some dot-like structures, which leads to dot-like atoms learned in the dictionary. Because such dot-like atoms are prone to match the noise in sparse representation process, the reconstructed image may contain some noises as a result. In addition, the noise distribution varies widely from inside the heart to surrounding tissues in low-dose thoracic CT, and severe noise usually appears around the cross-section of the vertebra region due to higher attenuation compared to other regions. Therefore, with a single dictionary and parameter setting, it is difficult to achieve optimal reconstruction everywhere within the whole image simultaneously.

In this paper, we propose the region-specific dictionary learning-based CT reconstruction method to address the above problem. Different regions such as the lungs and heart are separated by segmentation. We divide the image patches extracted from the whole image into four classes: the lung, heart, tissues, and boundary-juncture regions. Subsequently, the region-specific dictionaries are learned and used individually during image reconstruction. The proposed method takes advantage of that known of image features and noise of the system to adapt the difference among regions and thus reduces uncertainties in dictionary learning-based image reconstruction. For low-dose thoracic CT imaging, the region-specific dictionary can better recover anatomical details while removing the image noise effectively.

## II. METHODOLOGY

### A. Dictionary Learning-based CT Reconstruction

We first provide some background knowledge of the dictionary learning technique and the dictionary learning-based CT reconstruction [27, 31-33]. Dictionary learning usually works on small image patch. An image patch of pixels $W \times H$ can be expressed in vector form $X \in R^{N \times 1}$, $(N = W \times H)$. A dictionary is a matrix $D \in R^{N \times K}$, $(N \ll K)$, where each column is called an atom. In such a construct, if an image patch can be represented by the linear combination of only a few atoms of a dictionary with a small error tolerance ε, the sparse representation denoted as $\alpha \in R^{K \times 1}$, $(\|\alpha\|_0 \ll N)$ can be obtained by optimizing the problem in (1) or its unconstraint form in (2):

$$\min_{\alpha} \|\alpha\|_0, \ s.t. \ \|X - D\alpha\|_2^2 \leq \varepsilon \quad (1)$$

$$\min_{\alpha}(\|X - D\alpha\|_2^2 + \vartheta\|\alpha\|_0) \quad (2)$$

which can be solved by orthogonal matching pursuit (OMP) algorithm approximately [34].

Given a training set of patches $X_s \in R^{N \times 1}$, $(s = 1, \cdots, S)$, dictionary learning finds an atom set which can sparsely represent each patch of the training set, that is:

$$\min_{D, \alpha_s} \sum_s (\|X_s - D\alpha_s\|_2^2 + \vartheta_s \|\alpha_s\|_0) \quad (3)$$

Dictionary learning-based CT reconstruction aims to combine the dictionary-based sparse representation and the iterative CT reconstruction together so that the reconstructed image not only satisfies projection data fidelity, but also can be sparsely represented by the dictionary. Here the dictionary can be learned from a training set beforehand or intermediately from subsequent images during iteration. The former case is called global dictionary-based CT reconstruction and the latter is called adaptive dictionary-based CT reconstruction. Mathematically, dictionary learning-based CT reconstruction is to solve the following optimization problem [27]:

$$\min_{\mu,(D),\alpha_s} \|A\mu - p\|_w^2 + \lambda \sum_s (\|E_s\mu - D\alpha_s\|_2^2 + \vartheta_s \|\alpha_s\|_0) \quad (4)$$

where $\mu$ is the CT image, $p$ is the projection data, $A$ is the system matrix, $E_s$ is the patch extraction operator, and $\lambda$ is the regularization parameter. Equation (4) can be solved by the alternating minimization procedure described in [27].

In current dictionary learning-based CT reconstruction, whether global or adaptive, a single dictionary is used for the entire image. However, image structures and noise levels within the image may vary significantly from region to region. The "one size fits all" approach is therefore not optimal for CT image reconstruction, especially when large difference exist between subregions. Using region-specific dictionaries can give more accurate representations and lead to better reconstruction.

### B. Region-specific Dictionary Learning

For high performance reconstruction of low-dose CT, the dictionary should contain as many relevant features and as few irrelevant artifacts as possible. For thoracic CT images, the structural and noise distributions vary dramatically between the lungs, heart, and surrounding tissues. Therefore, we divide the image into different regions and reconstruct these regions by using region-specific dictionaries. In Fig. 1, we show an example of region segmentation and the corresponding patch classification. Therein, the red curves partition the whole thorax into the lungs, heart, and surrounding tissues according to the distinct structural and noise differences. For a patch of $\sqrt{N} \times \sqrt{N}$ pixels extracted from the image, if all the pixels belong to the surrounding tissue region (e.g. the blue blocks in Fig. 1), it is classified as the surrounding tissue patch. In this way, the patches belonging to the heart region (e.g. the purple blocks in Fig. 1) and the lung region (e.g. the green blocks in Fig. 1) can be obtained. A patch is classified as belonging to the juncture region if only some of the pixels belong to one region (e.g. the yellow blocks in Fig.1). By introducing the auxiliary juncture region, the lung, heart, and surrounding tissue regions can be unambiguously separated without stringent need for accurate segmentation of the regions. This is very useful in practice, especially in the low-dose situation. After the patches in each region are obtained, the dictionary can be learned from and used in each region, which is referred to as region-specific dictionary learning.

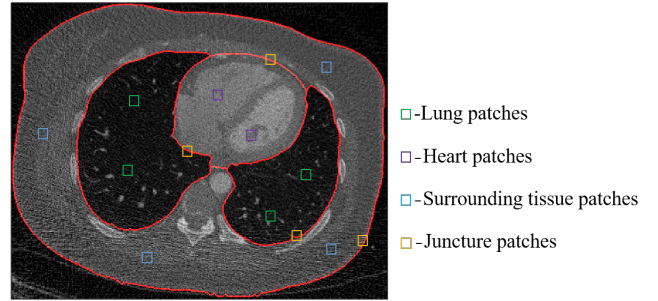

Fig. 1. Schematic diagram of region segmentation and patch classification. Juncture patch is the patch on the boundary of regions, where not all pixels belong to one region.

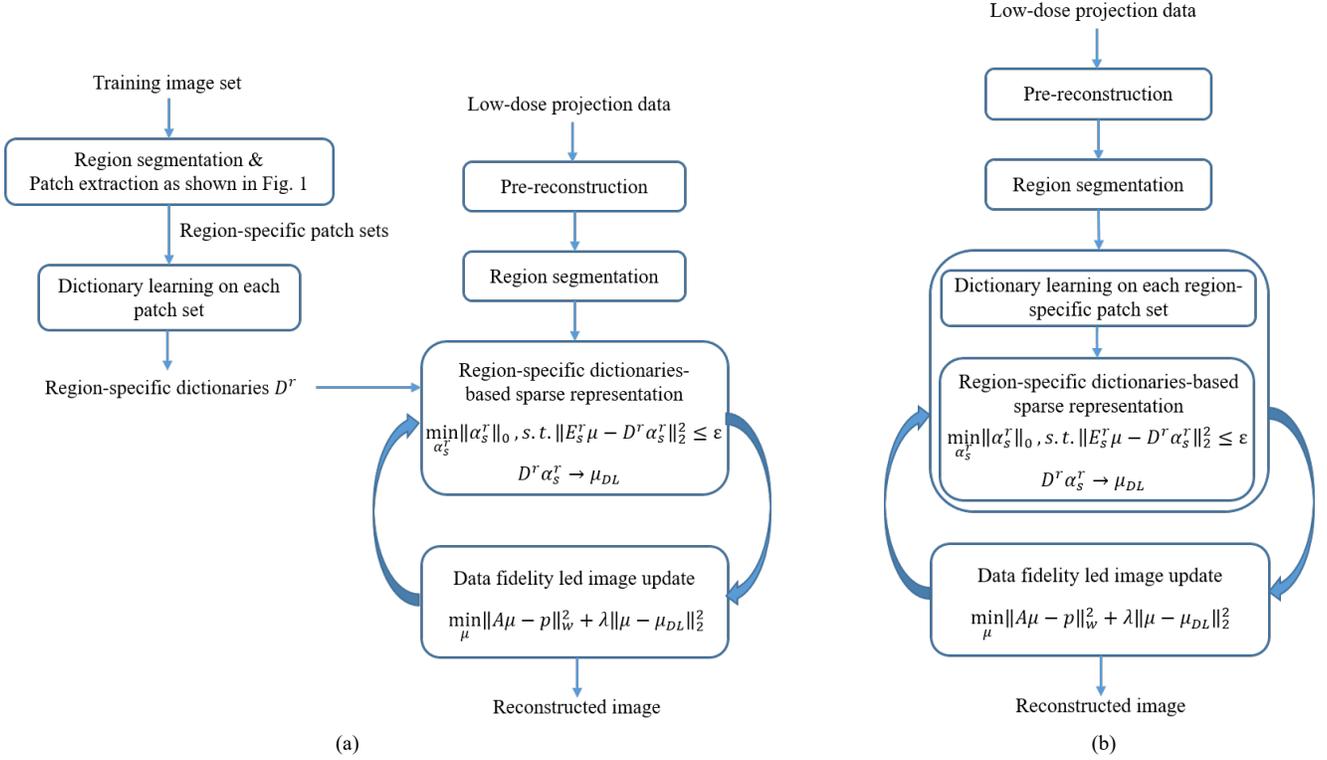

Fig. 2. Flow diagrams of the proposed methods. (a) illustrates steps in the region-specific global dictionary learning-based reconstruction, and (b) illustrates the region-specific adaptive dictionary learning-based reconstruction.

*C. Overall Workflow*

We recast (4) and write the proposed region-specific dictionary learning-based CT reconstruction as the optimization of following function:

$$\min_{\mu,(D^r),\alpha_s^r} \|A\mu - p\|_w^2 + \sum_r \lambda^r \sum_s (\|E_s^r\mu - D^r\alpha_s^r\|_2^2 + \vartheta_s^r \|\alpha_s^r\|_0) \quad (5)$$

where the superscript $r$ is the region index (i.e., the lung, heart, surrounding tissue, and juncture regions). Region-specific dictionaries are learned and used for each region. In this way, the dictionaries are more targeted and augment the recovery of local image structures while suppressing noise and/or artifacts.

Corresponding to the global and adaptive dictionary learning-based reconstruction frameworks, we present the proposed region-specific global and adaptive dictionary learning-based reconstruction workflows in Fig. 2. The pre-reconstruction is used to provide a basis for region determination and parameter selection. Usually, the FBP method is enough to satisfy the requirement. For region segmentation, the lung region is separated out because of its structural difference from other regions, whereas the heart region is separated out because of its big noise difference from surrounding tissues. The lung and heart regions are recognized automatically by some image segmentation methods or manually contouring [35-38]. In this paper, we simply combined the threshold based segmentation and region growing technique for the task of segmentation. For dictionary learning and sparse representation, we used the SPAMS toolbox [33, 39], which was found to be fast and effective. We followed the method described in [27] for parameter selection. It should be noticed that the sparse representation error, which is a very important parameter in the sparse representation step, should be selected according to noise level of each region. For the data fidelity-led image update step, the separable paraboloid surrogate method was utilized [40].

*D. Simulation and Experimental Evaluations*

In the numerical simulation, we selected ten normal-dose thoracic scans. One image was chosen to simulate a low-dose CT scan and the other nine images were used for dictionary learning. Image size was 360×440 pixels with pixel size of 0.075×0.075cm$^2$. Fan-beam equidistance geometry was simulated. The number of detector bins was 500. The diameter of the FOV was 51.73cm. Projections were simulated uniformly around 360°. Three kinds of projection numbers: 450, 225, and 150 were simulated. Poisson noise were simulated with three photon counts: 50000, 20000, and 10000. Hence, nine cases were simulated in total.

Besides the simulation study, we also evaluated the proposed method with human CT data, which was used in the study reported in [27]. Human thoracic CT sinograms of 64 slices were obtained by a GE CT750 HD scanner. Two sinograms were selected and down-sampled from 2200 views to 440 and 220 views to evaluate the performance on few-view data. The reconstruction size was 600×800 pixels with pixel size of 0.075×0.075cm$^2$ and the thorax region was 360×440 pixels.

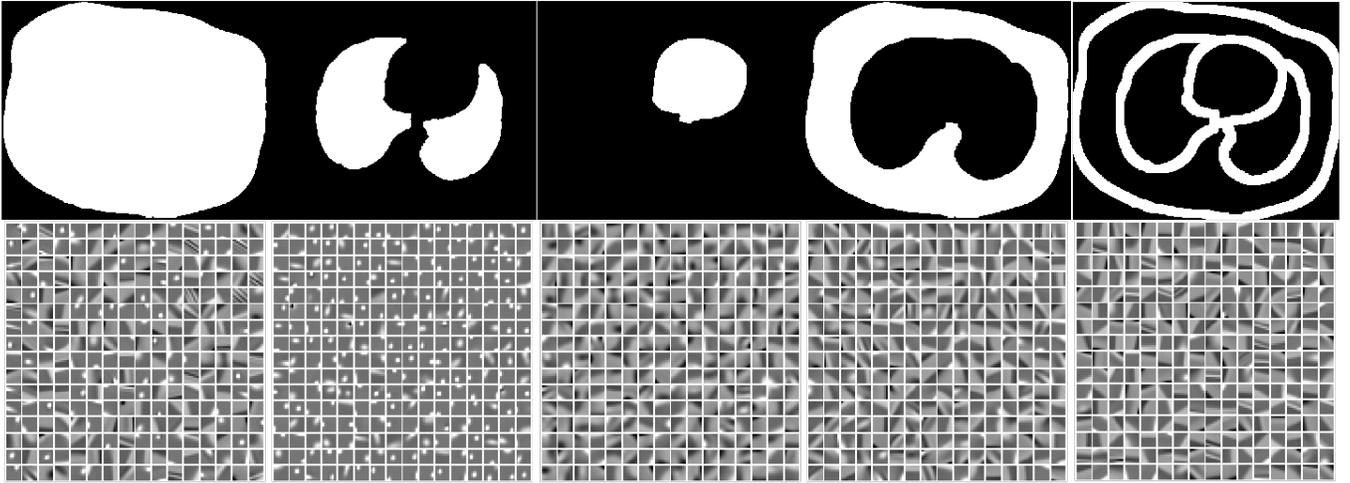

Fig. 3. Learned dictionaries (bottom) and their represented regions (top). The 1st column at left corresponds to the thorax region mask and represents the conventional single dictionary. The 2nd-5th columns are the lung, heart, surrounding tissue, and juncture region masks respectively, corresponding to the respective region-specific dictionaries.

For the described simulation and measurement data, the proposed Region-specific dictionaries and the conventional Single dictionary-based reconstructions under both the Global and Adaptive Dictionary learning-based reconstruction frameworks (abbr. as RGD, RAD, SGD, SAD respectively) were performed. The Structure-SIMilarity (SSIM) [41] and Root-Mean-Square Error (RMSE) were two indices used for quantitative comparisons. To demonstrate the improvement seen using the region-specific dictionary learning strategy, we introduce percent differences of the above two indices, which are defined in (6) and (7):

$$\Delta SSIM = \frac{\Delta SSIM_{GD} + \Delta SSIM_{AD}}{2} \times 100\%$$
$$= \left(\frac{SSIM_{RGD} - SSIM_{SGD}}{2 \times SSIM_{SGD}} + \frac{SSIM_{RAD} - SSIM_{SAD}}{2 \times SSIM_{SAD}}\right) \times 100\% \quad (6)$$

$$\Delta RMSE = \frac{\Delta RMSE_{GD} + \Delta RMSE_{AD}}{2} \times 100$$
$$= \left(\frac{RMSE_{SGD} - RMSE_{SGD}}{2 \times RMSE_{SGD}} + \frac{RMSE_{SAD} - RMSE_{SAD}}{2 \times RMSE_{SAD}}\right) \times 100\% \quad (7)$$

## III. RESULTS

### A. Construction of Region-specific Dictionaries

We selected nine normal-dose thoracic CT images for global dictionary learning. In the region-specific dictionary learning method, extracted patches were separated into four classes corresponding to the lung, heart, surrounding tissue, and juncture regions. Patch size was 8×8 pixels. The dictionary for each region had 256 atoms. The results are displayed in Fig. 3. It is seen that the lung dictionary has multiple dot-like atoms, which is very different from other dictionaries. The single dictionary learned from the whole thorax region aimed to express the features from all regions, thus it contained some dot-like atoms similar to those appearing in the dictionary specific to the lung. Atoms of the heart, surrounding tissue, and juncture dictionaries have similar appearance, which is consistent with their having similar structural characteristics.

### B. Simulation Results

Fig. 4 shows the reconstructed images of the lung, heart, and tissue regions using different numbers of simulated projections and photons. It is seen that dot-like noises appear in the surrounding tissue and heart regions in the conventional single dictionary-based reconstructions (SGD and SAD). With the use of region-specific dictionary learning-based reconstructions (RGD and RAD), such dot-like noise is effectively removed in all regions other than the lungs. At the same time, structures in lung region are seen more clearly with the proposed method.

We calculated SSIMs and RMSEs for the results reconstructed by the four methods. In Fig. 5, the results obtained using different numbers of simulated projections and photons are plotted. The results indicated that the region-specific method can improve the SSIM and decrease the RMSE effectively, especially in the low dose situation. In the low-dose situation, the global dictionary learning-based reconstruction outperforms the adaptive dictionary learning-based reconstruction due to high noise effects from dependence on only training with low-dose patches in the adaptive case, where normal-dose patches are also used in the global case.

TABLE I
RELATIVE DIFFERENCE PERCENTAGES OF THE SIMULATION RESULTS

| Region | Lungs | Heart | Surrounding tissues | Whole |
|---|---|---|---|---|
| ΔSSIM (%) | 0.58-4.17 | 0.53-4.09 | 0.75-5.41 | 0.68-4.88 |
| ΔRMSE (%) | 2.36-16.4 | 2.08-3.59 | 3.78-8.35 | 3.25-11.4 |

The relative difference percentages mentioned in (6) and (7) are shown in Table. 1. The values of these indices were calculated for each region and the entire thorax. Using the proposed strategy, the SSIMs of the entire thorax from nine simulated projection datasets were improved by at least 0.68%. In the lowest-dose case, a value as large as 4.88% was achieved. Meanwhile, the RMSEs were decreased by at least 3.25%. The largest value was found to be 11.4%. Specific to each region, the SSIMs of the nine simulated cases were improved by 0.58 to 4.17%, 0.53 to 4.09%, and 0.75 to 5.41% in the lung, heart, and the surrounding tissue regions, respectively. The corresponding decreases for the three regions in RMSEs are found as 2.36 to 16.4%, 2.08 to 3.59%, and 3.78 to 8.35%, respectively.

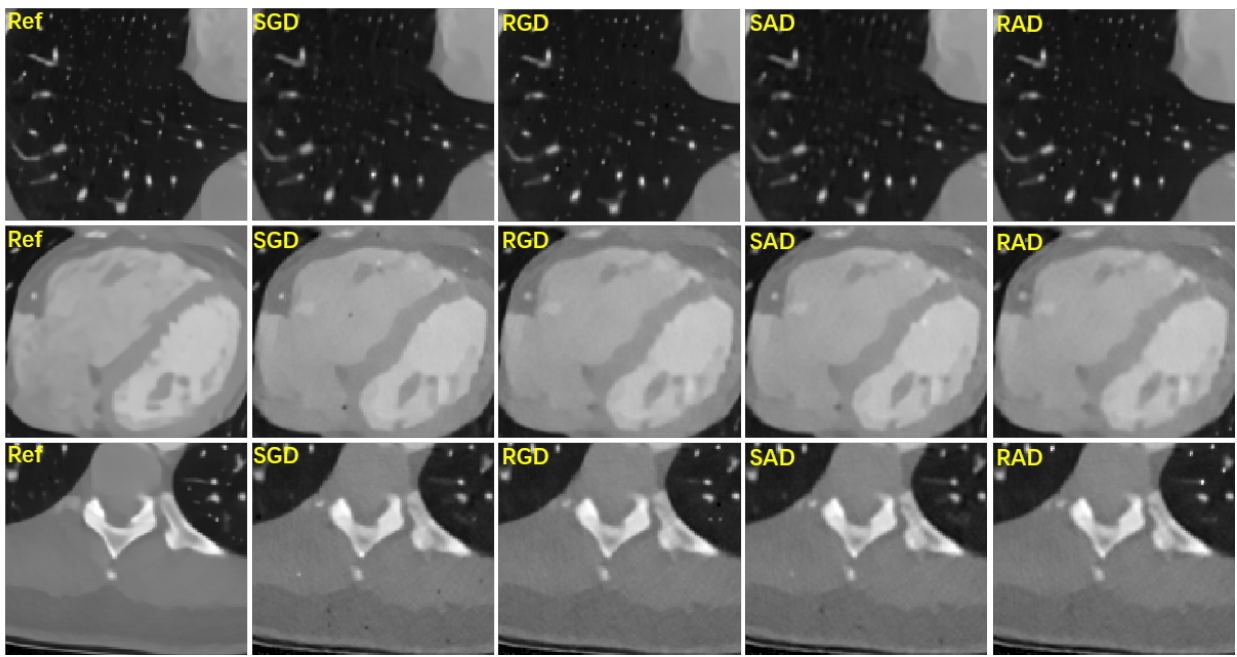

Fig. 4. Reconstruction results using simulation data. The 1st-3rd rows illustrate the results from data of 450 views with 50000 photons, 225 views with 20000 photons, and 150 views with 10000 photons, respectively. Display window is [0 0.4]/cm.

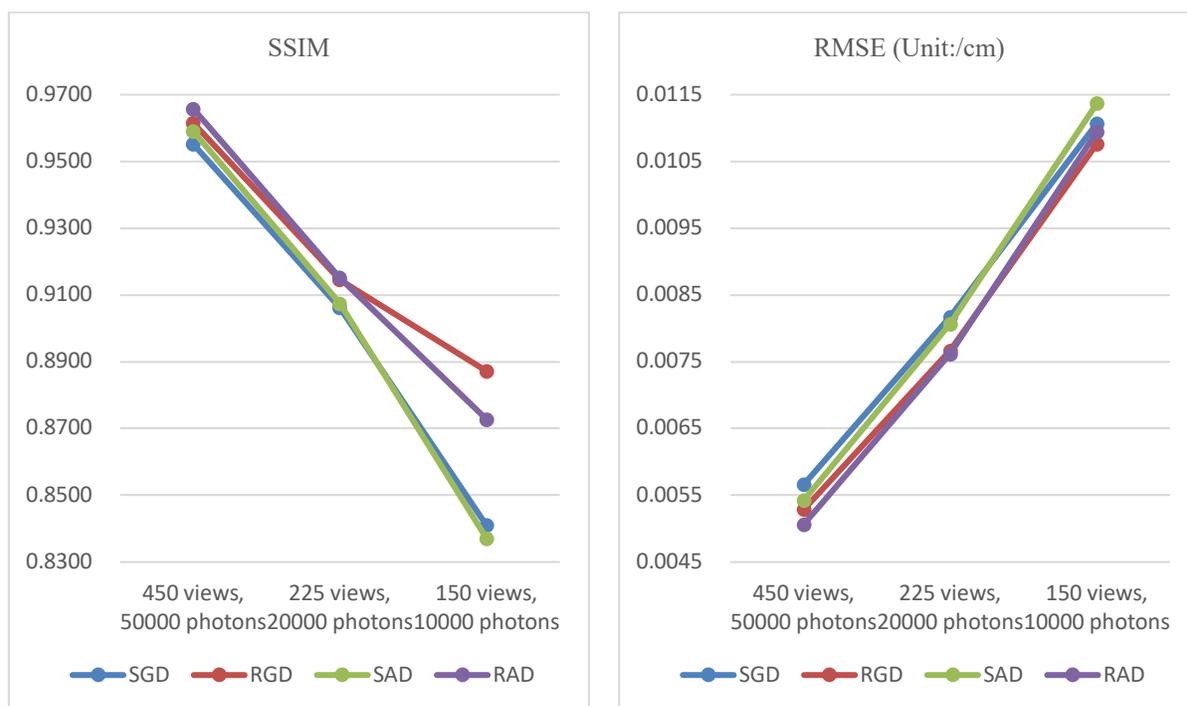

Fig. 5. Quantitative comparisons of the simulation results.

### C. Real-data Results

Reconstruction results based on sinograms with 440 and 220 views are shown in Fig. 6(a) and (b), respectively. In the single

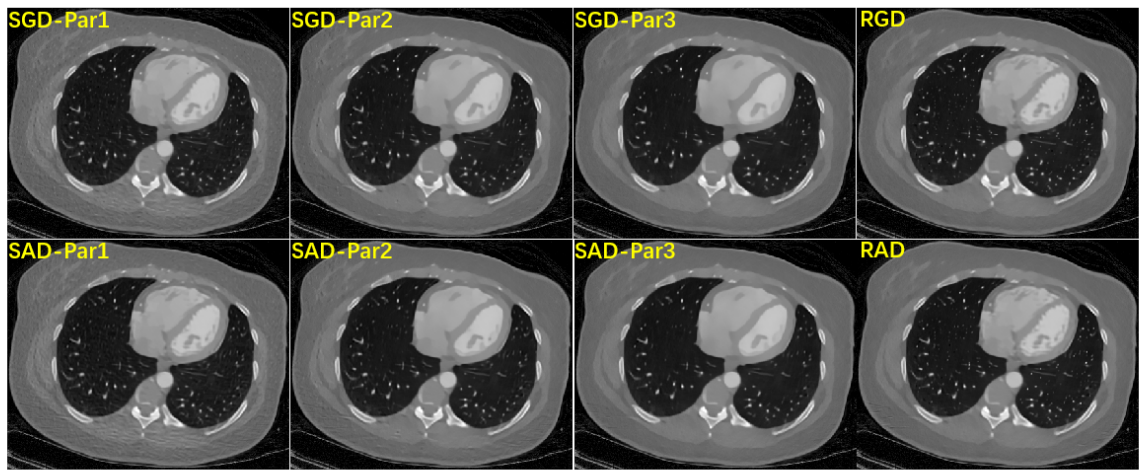

(a)

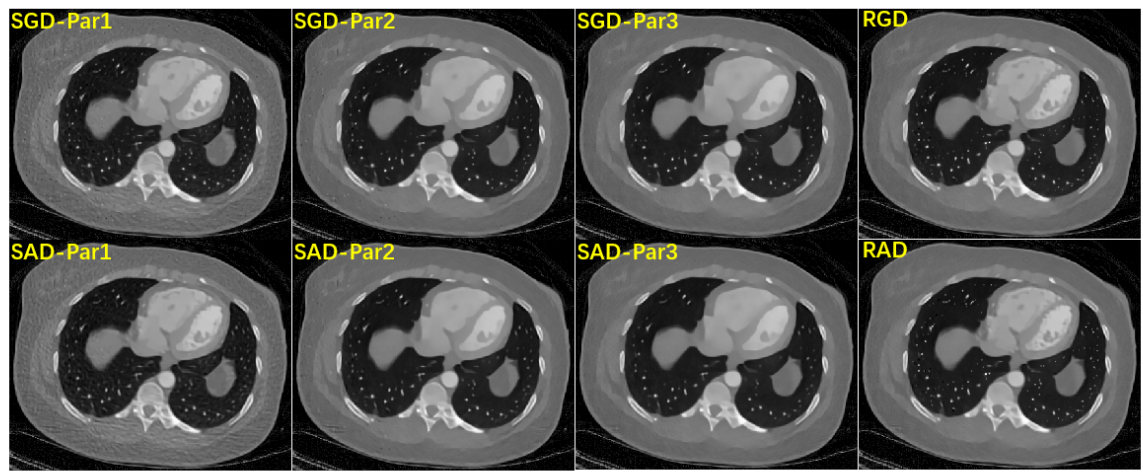

(b)

Fig. 6. Reconstruction results of real data. Par 1-3 indicate optimal parameterization for heart, lung, and surrounding tissue in the single dictionary method. (a) and (b) illustrate the results of one dataset with 440 projections and another dataset with 220 projections, respectively. Display window is [0 0.4]/cm.

TABLE II

QUANTITATIVE COMPARISONS BETWEEN SGD AND RGD METHODS. PAR 1-3 INDICATE OPTIMAL PARAMETERS FOR HEART, LUNG, AND SURROUNDING TISSUE IN SGD METHOD, RESPECTIVELY. (II-1) AND (II-2) ARE THE RESULTS FROM 440 VIEWS AND 220 VIEWS, RESPECTIVELY.

(II-1)

| Data with 440 views | | SGD | | | RGD | | |
|---|---|---|---|---|---|---|---|
| | | Par 1 | Par 2 | Par 3 | Par 1 | Par 2 | Par3 |
| SSIM | Lung | 0.9323 | **0.9445** | 0.9296 | 0.9356 | **0.9527** | 0.9423 |
| | Heart | **0.9813** | 0.9670 | 0.9536 | **0.9824** | 0.9704 | 0.9576 |
| | Body | 0.8993 | 0.9523 | **0.9671** | 0.9060 | 0.9587 | **0.9680** |
| | Whole | 0.9209 | 0.9520 | **0.9542** | 0.9239 | 0.9585 | **0.9590** |
| RMSE (/cm) | Lung | 0.0066 | **0.0062** | 0.0067 | 0.0060 | **0.0054** | 0.0059 |
| | Heart | **0.0035** | 0.0046 | 0.0057 | **0.0032** | 0.0042 | 0.0053 |
| | Body | 0.0064 | 0.0049 | **0.0043** | 0.0062 | 0.0045 | **0.0041** |
| | Whole | 0.0062 | 0.0053 | **0.0053** | 0.0058 | 0.0047 | **0.0048** |

(II-2)

| Data with 220 views | | SGD | | | RGD | | |
|---|---|---|---|---|---|---|---|
| | | Par 1 | Par 2 | Par 3 | Par 1 | Par 2 | Par3 |
| SSIM | Lung | 0.8745 | **0.9127** | 0.8943 | 0.8948 | **0.9248** | 0.9086 |
| | Heart | **0.9627** | 0.9445 | 0.9284 | **0.9652** | 0.9486 | 0.9337 |
| | Body | 0.8317 | 0.9315 | **0.9397** | 0.8411 | 0.9392 | **0.9420** |
| | Whole | 0.8658 | **0.9282** | 0.9262 | 0.8664 | **0.9366** | 0.9320 |
| RMSE (/cm) | Lung | 0.0085 | **0.0077** | 0.0086 | 0.0082 | **0.0072** | 0.0080 |
| | Heart | **0.0047** | 0.0059 | 0.0070 | **0.0045** | 0.0055 | 0.0066 |
| | Body | 0.0090 | 0.0064 | **0.0063** | 0.0086 | 0.0059 | **0.0058** |
| | Whole | 0.0083 | **0.0067** | 0.0071 | 0.0081 | **0.0062** | 0.0066 |

dictionary based reconstruction (SGD and SAD), multiple combinations of parameters were applied: 'Par 1', 'Par 2' and 'Par 3' in Fig. 6 indicate the optimal parameters found for the heart, lung, and surrounding tissue regions, respectively. It is seen that the surrounding tissue region is noisier than the central heart region. It is impossible to get optimal reconstruction in all regions simultaneously using the single dictionary method, whereas the region-specific dictionary can balance conflicting requirements of different regions effectively. The same conclusion can be drawn from quantitative comparisons (the gold standard being the results from the original data of 2200 views) shown in Fig. 7. The proposed strategy benefits from both the use of region-specific dictionaries and region-specific setting of optimization parameters. In Table II, we show the results of uniform parameter setting for SGD and RGD. We can see that, with parameterization mirrored from SGD, RGD achieves better image quality.

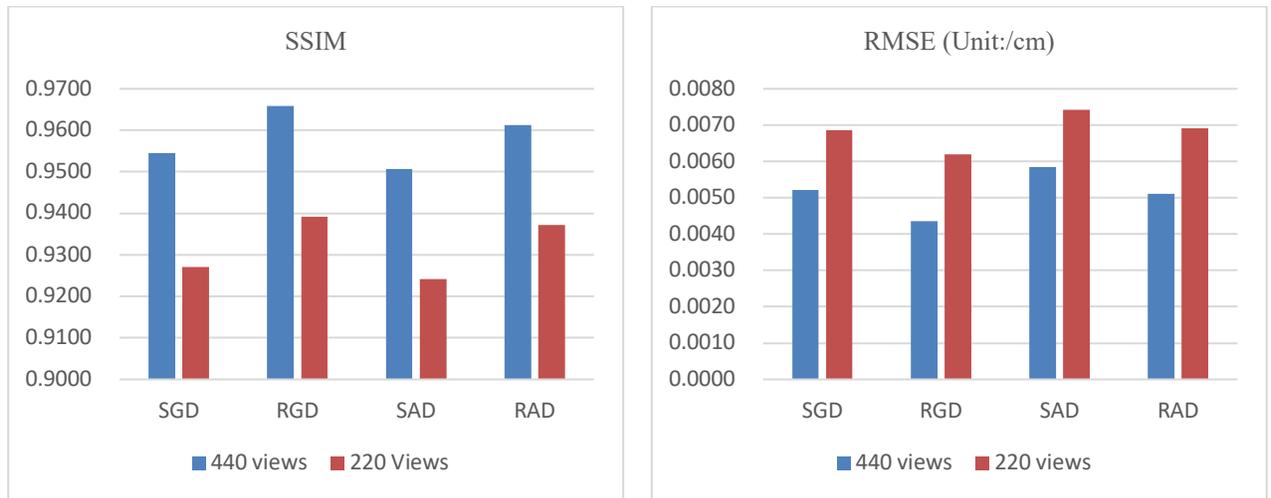

Fig.7. Quantitative comparisons of real data results.

## IV. Discussions

A region-specific dictionary learning strategy is presented to take consideration of the fact that the structural and noise distributions in thoracic CT are often region-specific. The utility of the method in improving the quality of low-dose thoracic CT reconstruction is demonstrated. We note that the results presented in last section depend on the selection of model parameters, which is similar to the many other image reconstruction methods. The results may change as different set of parameters are used. With careful selection of the reconstruction parameters in both single and region-specific dictionary learning, the results obtained by using the region-specific dictionary strategy seem to always outperform that of the single dictionary method. In that the single dictionary method can be considered a special case of the region-specific method, we introduce the latter as a generalization of the first.

In practice, the determination of regions for reconstruction is somewhat arbitrary. Structural and noise differences are two considerations employed in this work. We empirically segmented the lungs and heart from the thoracic CT based on the initial reconstruction using FBP. For segmentation, the threshold and region growing were used to delineate the lung and heart regions. However, more sophisticated methods can be applied for the task in future.

For dictionary learning and sparse representation, there are a few parameters to be determined. The selection of the patch size, the sparsity, and the representation error during sparse representation can affect the quality of resultant images. With the proposed method, the numbers of parameters increase with the numbers of regions. However, this does not increase the difficulty of parameter selection. In each region, we can simply follow the selection principle described in [27]. The region-specific strategy improves the goodness of the parameters through the manual classification of the atoms with coarse image segmentation. For example, the representation error parameter is calculated based on the noise level of the region, which is advantageous in terms of meeting the specifics of each region, instead of merely using a single uniform value for the entire image. In fact, even by simply following the parameter selection principle of the single dictionary method with a single parameter for the entire image, better results were seen (as shown in Table II) since the dictionary is regionally targeted. That said, parameter selection has long been recognized as a practical issue in any regularization-based reconstruction method and there exists an outstanding need for an effective technique of automated model parameter selection.

Computation time of the proposed region-specific method takes slightly longer as compared to the conventional single dictionary method due to the use of multiple dictionaries. The increase in computational burden is, however, insignificant as most computation time is spent on the forward-projection and back-projection in iterative reconstruction. The use of GPU-based parallel computing is one method to speed up the computation significantly[42].

Our work emphasizes the importance of considering heterogeneous characteristics of different anatomical regions in image reconstruction. The proposed region-specific strategy does not only apply to the dictionary learning-based CT reconstruction framework, but also other regularization-based iterative reconstructions and even analytical reconstruction methods. Structural and noise distributions are two representative features that can be used for the separation of regions. This kind of region-specific handling is essentially prior-knowledge based and useful to improve reconstruction in low-dose and/or other cases with insufficient data.

## V. Conclusion

In summary, we proposed the use of region-specific dictionaries for image reconstruction to address the disparate demands of different regions in thoracic CT. It is shown that the proposed method can better recover structural details while suppressing dot noise more effectively as compared to the conventional single dictionary method. The approach is particularly valuable in dealing with problems in which different regions have distinct image and noise features. For imaging problems in which the structural and noise distributions change little from region to region, the proposed method should perform better or at least equally well as the conventional single dictionary method.

## References

[1] D. P. Naidich, C. H. Marshall, C. Gribbin, R. S. Arams, and D. I. McCauley, "Low-dose CT of the lungs: preliminary observations," *Radiology,* vol. 175, pp. 729-731, 1990.

[2] H. Rusinek, D. P. Naidich, G. McGuinness, B. S. Leitman, D. I. McCauley, G. A. Krinsky*, et al.*, "Pulmonary nodule detection: low-dose versus conventional CT," *Radiology,* vol. 209, pp. 243-249, 1998.

[3] J. R. Mayo, J. Aldrich, and N. L. Müller, "Radiation exposure at chest CT: a statement of the Fleischner Society," *Radiology,* vol. 228, pp. 15-21, 2003.

[4] M. K. Kalra, M. M. Maher, T. L. Toth, L. M. Hamberg, M. A. Blake, J.-A. Shepard*, et al.*, "Strategies for CT radiation dose optimization," *Radiology,* vol. 230, pp. 619-628, 2004.

[5] H. Yu, S. Zhao, E. A. Hoffman, and G. Wang, "Ultra-low dose lung CT perfusion regularized by a previous scan," *Academic radiology,* vol. 16, pp. 363-373, 2009.

[6] T. Li, X. Li, J. Wang, J. Wen, H. Lu, J. Hsieh*, et al.*, "Nonlinear sinogram smoothing for low-dose X-ray CT," *IEEE Transactions on Nuclear Science,* vol. 51, pp. 2505-2513, 2004.

[7] P. J. La Riviere, "Penalized‐likelihood sinogram smoothing for low‐dose CT," *Medical physics,* vol. 32, pp. 1676-1683, 2005.

[8] J. Wang, T. Li, H. Lu, and Z. Liang, "Penalized weighted least-squares approach to sinogram noise reduction and image reconstruction for low-dose X-ray computed tomography," *IEEE transactions on medical imaging,* vol. 25, pp. 1272-1283, 2006.

[9] J. Wang, T. Li, Z. Liang, and L. Xing, "Dose reduction for kilovotage cone-beam computed tomography in radiation therapy," *Physics in medicine and biology,* vol. 53, p. 2897, 2008.

[10] T. Li, E. Schreibmann, B. Thorndyke, G. Tillman, A. Boyer, A. Koong*, et al.*, "Radiation dose reduction in four‐dimensional computed tomography," *Medical physics,* vol. 32, pp. 3650-3660, 2005.

[11] W. Xu and K. Mueller, "Efficient low‐dose CT artifact mitigation using an artifact‐matched prior scan," *Medical physics,* vol. 39, pp. 4748-4760, 2012.

[12] S. Li, H. Yin, and L. Fang, "Group-sparse representation with dictionary learning for medical image denoising and fusion," *IEEE Transactions on Biomedical Engineering,* vol. 59, pp. 3450-3459, 2012.

[13] Y. Chen, L. Shi, Q. Feng, J. Yang, H. Shu, L. Luo, *et al.*, "Artifact suppressed dictionary learning for low-dose CT image processing," *IEEE transactions on medical imaging,* vol. 33, pp. 2271-2292, 2014.
[14] H. Chen, Y. Zhang, W. Zhang, P. Liao, K. Li, J. Zhou, *et al.*, "Low-dose CT denoising with convolutional neural network," *arXiv preprint arXiv:1610.00321,* 2016.
[15] E. Kang, J. Min, and J. C. Ye, "A deep convolutional neural network using directional wavelets for low‐dose X‐ray CT reconstruction," *Medical physics,* vol. 44, 2017.
[16] E. Y. Sidky and X. Pan, "Image reconstruction in circular cone-beam computed tomography by constrained, total-variation minimization," *Physics in medicine and biology,* vol. 53, p. 4777, 2008.
[17] G. H. Chen, J. Tang, and S. Leng, "Prior image constrained compressed sensing (PICCS): a method to accurately reconstruct dynamic CT images from highly undersampled projection data sets," *Medical physics,* vol. 35, pp. 660-663, 2008.
[18] J. Wang, T. Li, and L. Xing, "Iterative image reconstruction for CBCT using edge‐preserving prior," *Medical physics,* vol. 36, pp. 252-260, 2009.
[19] H. Yu and G. Wang, "A soft-threshold filtering approach for reconstruction from a limited number of projections," *Physics in medicine and biology,* vol. 55, p. 3905, 2010.
[20] Y. Lu, J. Zhao, and G. Wang, "Few-view image reconstruction with dual dictionaries," *Physics in medicine and biology,* vol. 57, p. 173, 2011.
[21] L. Ritschl, F. Bergner, C. Fleischmann, and M. Kachelrieß, "Improved total variation-based CT image reconstruction applied to clinical data," *Physics in medicine and biology,* vol. 56, p. 1545, 2011.
[22] S. Singh, M. K. Kalra, M. D. Gilman, J. Hsieh, H. H. Pien, S. R. Digumarthy, *et al.*, "Adaptive statistical iterative reconstruction technique for radiation dose reduction in chest CT: a pilot study," *Radiology,* vol. 259, pp. 565-573, 2011.
[23] Q. Xu, X. Mou, G. Wang, J. Sieren, E. A. Hoffman, and H. Yu, "Statistical interior tomography," *IEEE transactions on medical imaging,* vol. 30, pp. 1116-1128, 2011.
[24] S. Baumueller, A. Winklehner, C. Karlo, R. Goetti, T. Flohr, E. W. Russi, *et al.*, "Low-dose CT of the lung: potential value of iterative reconstructions," *European radiology,* vol. 22, pp. 2597-2606, 2012.
[25] M. Katsura, I. Matsuda, M. Akahane, J. Sato, H. Akai, K. Yasaka, *et al.*, "Model-based iterative reconstruction technique for radiation dose reduction in chest CT: comparison with the adaptive statistical iterative reconstruction technique," *European radiology,* vol. 22, pp. 1613-1623, 2012.
[26] Y. Liu, J. Ma, Y. Fan, and Z. Liang, "Adaptive-weighted total variation minimization for sparse data toward low-dose x-ray computed tomography image reconstruction," *Physics in medicine and biology,* vol. 57, p. 7923, 2012.
[27] Q. Xu, H. Yu, X. Mou, L. Zhang, J. Hsieh, and G. Wang, "Low-dose X-ray CT reconstruction via dictionary learning," *IEEE Transactions on Medical Imaging,* vol. 31, pp. 1682-1697, 2012.
[28] H. Dang, J. W. Stayman, J. Xu, W. Zbijewski, A. Sisniega, M. Mow, *et al.*, "Task-based statistical image reconstruction for high-quality cone-beam CT," *Physics in Medicine & Biology,* vol. 62, p. 8693, 2017.
[29] Y. Zhang, X. Mou, G. Wang, and H. Yu, "Tensor-based dictionary learning for spectral CT reconstruction," *IEEE transactions on medical imaging,* vol. 36, pp. 142-154, 2017.
[30] H. Zhang, J. Ma, J. Wang, Y. Liu, H. Lu, and Z. Liang, "Statistical image reconstruction for low-dose CT using nonlocal means-based regularization," *Computerized Medical Imaging and Graphics,* vol. 38, pp. 423-435, 2014.
[31] M. Aharon, M. Elad, and A. Bruckstein, "k-SVD: An algorithm for designing overcomplete dictionaries for sparse representation," *IEEE Transactions on signal processing,* vol. 54, pp. 4311-4322, 2006.
[32] M. Elad and M. Aharon, "Image denoising via sparse and redundant representations over learned dictionaries," *IEEE Transactions on Image processing,* vol. 15, pp. 3736-3745, 2006.
[33] J. Mairal, F. Bach, J. Ponce, and G. Sapiro, "Online learning for matrix factorization and sparse coding," *Journal of Machine Learning Research,* vol. 11, pp. 19-60, 2010.
[34] J. A. Tropp and A. C. Gilbert, "Signal recovery from random measurements via orthogonal matching pursuit," *IEEE Transactions on information theory,* vol. 53, pp. 4655-4666, 2007.
[35] M. S. Brown, M. F. Mcnitt-Gray, N. J. Mankovich, J. G. Goldin, J. Hiller, L. S. Wilson, *et al.*, "Method for segmenting chest CT image data using an anatomical model: preliminary results," *IEEE transactions on medical imaging,* vol. 16, pp. 828-839, 1997.
[36] S. Hu, E. A. Hoffman, and J. M. Reinhardt, "Automatic lung segmentation for accurate quantitation of volumetric X-ray CT images," *IEEE transactions on medical imaging,* vol. 20, pp. 490-498, 2001.
[37] O. Ecabert, J. Peters, H. Schramm, C. Lorenz, J. von Berg, M. J. Walker, *et al.*, "Automatic model-based segmentation of the heart in CT images," *IEEE transactions on medical imaging,* vol. 27, pp. 1189-1201, 2008.
[38] J. Larrey-Ruiz, J. Morales-Sánchez, M. C. Bastida-Jumilla, R. M. Menchón-Lara, R. Verdú-Monedero, and J. L. Sancho-Gómez, "Automatic image-based segmentation of the heart from CT scans," *EURASIP Journal on Image and Video Processing,* vol. 2014, p. 52, 2014.
[39] http://spams-devel.gforge.inria.fr/.
[40] I. A. Elbakri and J. A. Fessler, "Statistical image reconstruction for polyenergetic X-ray computed tomography," *IEEE transactions on medical imaging,* vol. 21, pp. 89-99, 2002.
[41] Z. Wang, A. C. Bovik, H. R. Sheikh, and E. P. Simoncelli, "Image quality assessment: from error visibility to structural similarity," *IEEE transactions on image processing,* vol. 13, pp. 600-612, 2004.
[42] G. Pratx and L. Xing, "GPU computing in medical physics: A review," *Medical physics,* vol. 38, pp. 2685-2697, 2011.